\documentclass[manuscript]{aastex}

\slugcomment{Accepted for publication in ApJ, 03/11/14}

\shorttitle{Core-Halo Age Gradients}
\shortauthors{Getman et al.}

\begin{document}

\title{Core-Halo Age Gradients and Star Formation in the Orion Nebula and NGC~2024 Young Stellar Clusters}

\author{Konstantin V. Getman, Eric D. Feigelson, Michael A. Kuhn}

\affil{Department of Astronomy \& Astrophysics, 525 Davey Laboratory, Pennsylvania State University, University Park, PA 16802, USA}

\begin{abstract}
We analyze age distributions of two nearby rich stellar clusters, the NGC 2024 (Flame Nebula) and Orion Nebula Cluster (ONC) in the Orion molecular cloud complex. Our analysis is based on samples from the MYStIX survey and a new estimator of pre-main sequence (PMS) stellar ages, $Age_{JX}$, derived from X-ray and near-infrared photometric data. To overcome the problem of uncertain individual ages and large spreads of age distributions for entire clusters, we compute median ages and their confidence intervals of stellar samples within annular subregions of the clusters. We find core-halo age gradients in both the NGC 2024 cluster and ONC: PMS stars in cluster cores appear younger and thus were formed later than PMS stars in cluster peripheries. These findings are further supported by the spatial gradients in the disk fraction and $K$-band excess frequency. Our age analysis is based on $Age_{JX}$ estimates for PMS stars, and is independent of any consideration of OB stars. The result has important implications for the formation of young stellar clusters.  One basic implication is that clusters form slowly and the apparent age spreads in young stellar clusters, which are often controversial, are (at least in part) real. The result further implies that simple models where clusters form inside-out are incorrect, and more complex models are needed. We provide several star formation scenarios that alone or in combination may lead to the observed core-halo age gradients.
\end{abstract}


\section{Introduction \label{intro_section}}

While considerable insights have been gained regarding the formation of stars on small scales, the formation of rich stellar clusters dominated by OB stars is quite uncertain.  Debate has waged for decades over the relative importance of rapid  `top-down' fragmentation or a slower `bottom-up' process involving merging of subcomponents  \citep{Clarke00}.   In the former model, cluster formation would occurring within $1-2$ core free-fall times, $t_{ff} \simeq 10^5$~yr \citep{Elmegreen00}.  Rapid gravitational collapse is characteristic of clouds with weak turbulence on small scales \citep[]{VazquezSemadeni03}.  In the latter model, cluster formation can be extended over millions of years, delayed by strong turbulence in the natal cloud.  \citet{Krumholz07} argue for slow cluster formation, in part on extensive evidence for strong turbulence in giant molecular clouds \citep{MacLow04} and in part on empirical evidence from Hertzsprung-Russell diagrams (HRDs) of revealed clusters showing a wide age spread in the apparent ages of pre-main sequence (PMS) stars. 

However, the interpretation of HRD age spreads is quite controversial for both observational and theoretical reasons, as reviewed by \citet{Preibisch12}.   For example, \citet{Huff06} construct a model of the Orion Nebula Cluster (ONC) where star formation accelerates slowly over millions of years in both the cluster core and outer regions.  From a detailed analysis based on optical photometry and spectroscopy, \citet{Reggiani2011} report a characteristic age of 2.2~Myr, and an intrinsic age spread of 2~Myr, after accounting for various sources of age uncertainties. \citet{Jeffries2011} argue that the similar apparent age distribution of disk-bearing and disk-free stars in the ONC implies that the cluster does not have a wide intrinsic age spread, but their result permit a spread of $\sim 3$~Myr.  A variety of observational difficulties can also cause artificial age spreads \citep{Preibisch12}.  And astrophysical calculations of stars decending the Hayashi tracks for different histories of accretion show that an artificial age spread can appear for coeval clusters under certain conditions \citep{Baraffe12}. 

A confusing element of this issue is that clusters often \citep[though not always;][]{Wang08} show mass segregation where massive OB stars concentrate in the cluster cores.  The ages of massive stars are difficult to estimate as they quickly arrive on the main sequence while the PMS stars are on their Hayashi tracks.  There is some evidence that massive stars in cluster cores are younger than more dispersed PMS stars in rich clusters (\S\ref{prev_evidence.section}).  But it is not clear whether this reflects a general age gradient in cluster formation, or some special slow mode in the star formation process of massive stars, such as competitive accretion of core gas or stellar mergers \citep{Zinnecker07}. 

In this context, we present a new empirical finding: that PMS stars are younger in the core region than in the outer regions of two nearby rich clusters, the NGC~2024 (Flame Nebula) and Orion Nebula clusters in the Orion molecular cloud complex. This result is independent of the ages of massive stars in the core.   The work rests on two foundations.  The first is a new sample of cluster members developed by the MYStIX (Massive Young Star-Forming Complex Study in Infrared and X-ray) project for 20 star forming regions, including the Flame and Orion Nebulae \citep{Feigelson13}. The second is a new estimator of PMS stellar ages called $Age_{JX}$ that is derived from X-ray and near-infrared photometric data \citep[][hereafter G14]{Getman2014}.   The spatial-age gradient we report here lies within the unified structure of a rich monolithic cluster on $< 1$~pc scales.  Getman et al.\ report a separate result of spatial-age gradients across star forming regions on $\sim 2-20$~pc scales.

We briefly review the methodology and samples from the MYStIX and $Age_{JX}$ studies (\S\ref{methods.section}), and present the results for the NGC~2024 and Orion Nebula clusters in \S\ref{results_section}.  The implications for rich cluster formation are discussed in \S\ref{discussion.section}.

\section{Methods and Samples} \label{methods.section}

The MYStIX project has produced rich stellar samples of 31,784 probable members of 20 OB-dominated young star forming regions at distances $0.4-3.6$~kpc using sensitive X-ray, near-infrared (NIR), and mid-infrared photometric catalogs \citep{Feigelson13}. Based on the MYStIX data, G14 developed a new PMS stellar age estimator, $Age_{JX}$. This estimator is based on $J$-band stellar photospheric emission and on X-ray emission from coronal magnetic flaring, and relies on an assumption of a nearly universal $L_X-M$ relationship for $\la 5$~Myr old PMS stars (\S 3.1 in G14)\footnote{Similar $L_X - M$ relationships have been seen in a number of star forming regions with different ranges of characteristic ages, such as  W~40 ($\la 1$~Myr), Taurus and ONC ($1-2$~Myr), and IC~348, NGC~2264, and Cep~OB3b ($2- \ga 3$~Myr).}. Stellar masses are directly derived from absorption-corrected X-ray luminosities calibrated to the $L_X - M$ relation from the Taurus cloud PMS population. These masses are combined with $J$-band magnitudes, corrected for source extinction (using the NIR color-color diagram; \S 2.3 in G14) and distance, for comparison with PMS evolutionary models \citep{Siess2000} to estimate ages.

$Age_{JX}$ estimates are computed for over 5500 MYStIX PMS stars in 20 star forming regions (G14). However, as for any other age method, $Age_{JX}$ estimates for individual stars are highly uncertain. The largest contribution to the error on $Age_{JX}$ is the scatter on the $L_X-M$ diagram that causes typical individual $Age_{JX}$ uncertainties of $\sim 2$~Myr (\S 3.3 in G14). To overcome the problem of uncertain individual ages and large spreads of age distributions for entire MYStIX regions, G14 compute median ages and their confidence intervals of stellar samples within individual MYStIX subclusters, which in turn were identified through the spatial clustering analysis by \citet{Kuhn2014}. This approach allows to discriminate age differences and gradients among subclusters. The $Age_{JX}$ estimates for over a hundred MYStIX subclusters are referenced to a uniform time scale, which depends on the \citet{Siess2000} PMS evolutionary models. Since the $Age_{JX}$ method relies on evolutionary models, it does not provide ``absolute'' ages, as with most the standard PMS age methods. 

In this work, a similar approach is applied to the stars within the NGC 2024 and ONC clusters. Calculating median ages and their confidence intervals for different subregions of interest within these clusters, allowed the discovery of cluster core-halo age gradients (\S \ref{age_gradient_section}). 

We use subsamples of MYStIX PMS stars with available $Age_{JX}$ estimates. The stars have been stratified by elliptical annular subregions within the clusters. Figures~\ref{fig1} (panels a and d) show these annuli with the inner and outer ellipses encompassing 1 and $\ga 10$ cluster core radii, respectively. The NGC 2024 cluster core ellipses are adapted from \citet{Kuhn2014}. The ONC cluster core ellipses are calculated here using the spatial structure analysis methodology of \citet{Kuhn2014}. 

For NGC~2024, the original $Age_{JX}$ sample of G14 is relatively small (103 stars), with only 4 stars in the cluster core. We therefore relax their conservative $Age_{JX}$ selection criterion on the reliability of NIR data ($\sigma_{JHK_s} < 0.1$~mag) to add additional $\sim 20$ sources, all with their $JHK_s$ errors slightly above $0.1$~mag. This increases the number of $Age_{JX}$ sources in the cluster core to 9 stars and brings the total number of NGC~2024-$Age_{JX}$ stars to 121.  The NGC~2024 core cluster ellipse is centered at $(\alpha,\delta) = (5^h41^m42.5^s, -1^\circ54\arcmin13.7\arcsec)$ (J2000) with core radius (average of the major and minor axes) $r_c = 0.13$~pc, ellipticity $\epsilon = 0.37$, and position angle $\phi = 146^\circ$ east of north \citep{Kuhn2014}. The size of these annular subsamples are 9, 22, 66, and 24 stars from the cluster core towards the halo. The boundaries of the annular regions in units of the cluster core radius are $[0-1]$ (red ellipse in Figure~\ref{fig1}a), $[1-2]$ (green), $[2-6]$ (blue), and $[6-10]$ (orange).  

For the ONC, the MYStIX-Orion $Age_{JX}$ sample of G14 is truncated to a subsample of 391 lightly-absorbed stars ($J-H<1.2$~mag corresponding to $A_V \la 5$~mag). This reduces `contamination' by Orion Molecular Cloud members, including deeply embedded protostars in the Becklin-Neugebauer/Kleinman-Low and OMC-1S star formation regions \citep{Grosso2005}. The ONC core cluster ellipse is centered at $(\alpha,\delta) = (5^h35^m16.8^s,-5^\circ23\arcmin22.4\arcsec)$ (J2000) with average $R_{core}=0.12$~pc, $\epsilon = 0.47$, and $\phi = 176^\circ$.  The annular regions are $[0-1]$ (red ellipse in Figure~\ref{fig1}d), $[1-2]$ (green), $[2-5]$ (blue), $[5-7]$ (orange), and $[7-15]$ (cyan) core radii. The sample sizes in the annuli are 55, 57, 109, 83, and 87 stars from the center to the halo, respectively.  

Table \ref{tbl_stars} presents the resulting NGC~2024 and ONC $Age_{JX}$ catalogs including star position, angular distance from the cluster center, X-ray luminosity and age estimates, the flag for the presence of a circumstellar disk, and the flag for the membership in different cluster annular subregions.

The color points in Figure~\ref{fig1} (panels c and f) are median age values for stars in each annular subregion. The error bars are 68\% confidence intervals of the medians obtained from the bootstrap resampling of the stellar subsamples (G14).  Figure~\ref{fig1} (panels b and e) show B-spline fits \citep{HeNg1999} for the 25\%, 50\% (median), and 75\% quartiles to the $Age_{JX}$ values as a function of angular distance from the centers of the clusters.  This spline quantile least squares regression method is described by \citet{HeNg1999} and implemented in the R statistical software system in package $CABS$ \citep{Ng11}. These methods were used in G14, except the  spline regression fits local parabolic rather than linear functions.

\section{Results \label{results_section}}

\subsection{Discovered Age Gradients \label{age_gradient_section}}

The median spline fits in Figure~\ref{fig1} (panels b and e) show monotonic increases in age within 1~pc from the centers to the halos of both clusters: from $\sim 0.1$~Myr to $\sim 1.8$~Myr in NGC~2024, and from $\sim 1.4$~Myr to $\sim 2$~Myr in the ONC. The large observed age spreads of individual stars about the median with the typical interquartile ranges $\sim 1$~Myr and $\sim 2$~Myr in NGC~2024 and ONC, respectively, are a consequence of uncertainties of individual ages, possibly combined with real age spreads.  

But the median ages in the annuli (Figure~\ref{fig1} panels c and f) are quite well-constrained. For NGC~2024, the stars in the cluster core ($R < 0.15$~pc) are significantly younger with median age $0.2 \pm 0.1$~Myr than the stars in the halo, $1.1 \pm 0.2$ and $1.5 \pm 0.3$~Myr in the $0.26 < R < 0.78$~pc and $0.78 < R \la 1$~pc annuli, respectively. For the ONC, the age gradient is apparent only in the innermost annulus.  The median age of stars in the cluster core ($R < 0.15$~pc) is $1.2 \pm 0.2$~Myr compared to $1.9 \pm 0.1$ for the combined outer annuli with $0.15 < R \la 1$~pc.  These age gradients have very high statistical significance ($p << 0.001$).  

This gradient in stellar ages is associated with a gradient in interstellar absorption in the expected sense that younger PMS stars are typically more heavily absorbed.  In NGC~2024, source extinction monotonically decreases from the core (median $A_V \sim 20$~mag)  to the periphery (median $A_V \sim 6$~mag) of the cluster\footnote{For the NGC 2024 subregions the median $J-H$ colors [and $A_V$ estimates] are: $2.95\pm0.24$~mag [$19.6\pm2.2$~mag] (red ellipse in Figure~\ref{fig1}), $1.73 \pm 0.18$~mag [$10.0\pm1.1$~mag] (green), $1.60 \pm 0.08$~mag [$8.4\pm0.7$~mag] (blue), and $1.29 \pm 0.08$~mag [$5.7\pm0.9$~mag] (orange).}. For the ONC, the stars within the core (red) and the 2 inner halo subregions (green and blue) are slightly more absorbed (median $A_V \sim 2$~mag) than the stars within the outer halo (orange and cyan; median $A_V \sim 1.4$~mag), but this may be due to gradients in the obscuring veil of material that lies in front of the ONC \footnote{For the ONC subregions the median $J-H$ colors [and $A_V$ estimates] are: $0.95\pm0.02$~mag [$2.1\pm0.2$~mag] (red ellipse in Fig.~\ref{fig1}), $0.93 \pm 0.02$~mag [$1.8\pm0.3$~mag] (green), $0.95 \pm 0.03$~mag [$2.1\pm0.2$~mag] (blue), $0.83 \pm 0.03$~mag [$1.4\pm0.3$~mag] (orange), and $0.82 \pm 0.02$~mag [$1.4\pm0.2$~mag] (cyan).}. 

Lastly, we extend the ONC age analysis to a larger stellar sample with higher extinctions up to $A_{V,lim} = 30$~mag, thereby increasing contamination from young stellar objects embedded in the OMC cloud.  The resulting median ages are  younger than in the lightly obscured sample: from $1.8$~Myr to $1.5$~Myr in the green and blue elliptical annuli, and from $2$~Myr to $1.7$~Myr in the orange annulus.   The median ages in the core (red ellipse) and in the most outward halo region (cyan) are unchanged at $1.2$~Myr and $1.9$~Myr, respectively. These two regions have the fewest contaminating OMC objects.

\subsection{Mitigating Biases and Data Selection Effects \label{xlf_section}}

The MYStIX samples are mostly constructed as the union of carefully defined samples at three wavebands with some advanced statistical processing, applied uniformly to 20 massive star forming regions \citep[see Figure 3 in][]{Feigelson13}.  The constituent samples are from the {\it Chandra X-ray Observatory}, the {\it Spitzer Space Telescope}, and 2MASS and UKIRT/UKIDSS NIR surveys. However, various MYStIX data selection and incompleteness biases are present (see Appendix of Feigelson et al.) and could potentially affect our age gradient results.


\subsubsection{Possible Biases \label{xlf_biases_subsubsection}}

Ideally, it is desirable to perform analyses of cluster age gradients using spatial star samples with similar completeness limits in mass because: (a) the real age distribution of a stellar cluster could be astrophysically mass-dependent; and (b) the location of data and evolutionary model isochrones on the HRD or a CMD diagram (or the $M_J-M$ diagram in case of $Age_{JX}$) could be inconsistent with each other for different mass ranges. However, the following possible incompleteness and selection biases in the MYStIX data of NGC 2024 and ONC might complicate the construction of uniform samples complete to some mass limit. 

\begin{description}

\item First, we use subsamples of MYStIX PMS stars with available $Age_{JX}$ estimates (\S \ref{methods.section} and G14). For the $Age_{JX}$ analysis, the full MYStIX PMS samples were restricted to low-mass stars with available $L_X$ measurements with $10^{29} < L_X < 3 \times 10^{30}$~erg s$^{-1}$ ($0.2 \la M \la 1.2$~M$_{\odot}$) and reliable NIR photometry,  and were culled of protostellar objects with extremely large $K_s$-excesses for which there is no clear NIR dereddening procedures (G14). 

\item Second, the sensitivity to PMS stars in the X-ray data has a spatial bias due to the decrease in the {\it Chandra} telescope sensitivity with increasing off-axis angle, and due to the MYStIX bias against highly dispersed PMS stars. 

\item Third, the reduced sensitivity to ONC members in the {\it Chandra}-ACIS image of Orion within $1\arcmin$ of $\theta^{1}$~Ori~C due to the extended wings of its point spread function.

\item Forth, the limiting sensitivity to faint X-ray sources in NGC 2024 due to the shorter exposure time of its {\it Chandra} observation \citep[Table 2 in][]{Feigelson13}.

\item Fifth, The reduced sensitivity to NIR and X-ray NGC 2024 members in the central region due to the presence of an obscuring molecular core. 

\item Sixth, the reduced sensitivity to faint NIR stars due to the presence of high background nebular emission from heated dust around the cluster center. 

\end{description}

\subsubsection{ONC \label{xlf_onc_subsection}}

For the analysis of age gradients in ONC, the six potential incompleteness and selection biases described in \S \ref{xlf_biases_subsubsection} were easily overcome due to the exceptionally deep {\it Chandra}-ACIS exposure of the Orion region, nicknamed the $Chandra$ Orion Ultradeep Project \citep[COUP;][]{Getman05}.  The ``COUP unobscured'' stellar sample of 839 cool ONC stars \citep{Feigelson05} was found to be $\sim 90$\% complete down to $M \sim 0.2$~M$_{\odot}$ (neglecting effects of binarity). This is shown in Figure 19 of \citet{Getman06} that compares the initial mass function (IMF) of the ``COUP unobscured'' sample with the ONC IMF from \citet{Muench02}.
The $Age_{JX}$ ONC subsample used in our age gradient analyses is identical to the ``COUP unobscured'' stars with the constraint $10^{29} < L_X < 3 \times 10^{30}$~erg s$^{-1}$ applied.  

The first two potential sample bias problems are thus not present in the ONC sample.  The third problem is mitigated by the  $L_X$ constraint that removes all faint X-ray sources from the sample.  The fourth and fifth problems are not relevant to the ONC, while the sixth potential problem was overcome by correlating X-ray sources with a deep NIR image obtained with the VLT/ISAAC instrument on an 8-meter telescope \citep{Getman05}.  Thus, we expect that the observed mass function of the $Age_{JX}$ ONC stellar sample is representative of the IMF for $\sim 90$\% of the intrinsic ONC population within the mass range between $\sim 0.2$~M$_{\odot}$ and $1.2$~M$_{\odot}$, if effects of binarity are ignored. 

Using the X-ray luminosity function (XLF) as a surrogate for the IMF \citep{Getman06}, we investigate whether the X-ray luminosities have a spatial gradient that may have artificially produced an age gradient through the $L_X-M$ (X-ray-mass) relationship central to the $Age_{JX}$ estimation method (G14). Figure~\ref{fig_xlfs}c shows the cumulative XLFs  for the core-halo strata in the ONC. The strata have statistically indistinguishable XLFs, as confirmed with Kolmogorov-Smirnov (KS) tests. The KS test probabilities between the red and the remaining strata are $P_{KS} = 95$\% (green), $P_{KS} = 94$\% (blue), $P_{KS} = 28$\% (orange), and $P_{KS} = 95$\% (cyan). Since the $L_X-M$ relation remains unvarying with time during the early evolutionary phase of PMS stars (G14) and the XLF is a surrogate for IMF, the mass functions of the $Age_{JX}$ stellar subsamples within the different core-halo subregions are expected to be similar as well.

\subsubsection{NGC 2024 \label{xlf_ngc2024_subsection}}

In the case of NGC~2024, the mutual effect of the first ($Age_JX$ sample selection), fourth (moderate X-ray exposure), and especially the fifth (high source extinction from the molecular core) and sixth (high background NIR nebula emission) biases (\S \ref{xlf_biases_subsubsection}) is difficult to overcome, so the resulted $Age_{JX}$ stellar subsamples are more heterogeneous than in the ONC.  

All spatial $Age_JX$ strata in NGC~2024 have statistically indistinguishable observed XLFs (Figure~\ref{fig_xlfs}a), as confirmed with KS tests\footnote{For instance, the KS test probabilities between the orange and the remaining strata are $P_{KS} = 74$\% (red), $P_{KS} = 99$\% (green), and $P_{KS} = 11$\% (blue). The KS test probabilities between the red and the remaining strata are $P_{KS} = 63$\% (green), $P_{KS} = 19$\% (blue), and $P_{KS} = 74$\% (orange).}. However, as a result of the first and fourth potential biases, the three $Age_{JX}$ source subsamples for the NGC~2024 halo strata (green, blue, orange) are $\sim 80-90$\% complete down to only $\log(L_X) \sim 30$~erg/s ($0.6$~M$_{\odot}$) when compared to the XLF of the ``COUP unobscured'' population.  The problems are particularly acute in the cluster core where -- due to the mutual effect of the first, forth, fifth, and sixth biases --  the respective $Age_{JX}$ subsample is incomplete for any range of $L_X$ or mass. A handful of $Age_{JX}$ sources in the core of NGC~2024 thus might not be representative of the entire stellar population in the cluster core.  We now provide additional lines of evidence for the extreme youth of the cluster core in NGC~2024.

\subsection{Age of the NGC~2024 Core \label{ngc2024_core_section}}

In the NGC~2024 cluster core, out of 114 MPCMs \citep{Broos2013, Kuhn2014, Povich2013}, 95 have X-ray column density estimates $N_H$, $34$ have disk classifications based on NIR and $Spitzer$ photometry, $22$ have $JHK_s$ measurements (not upper limits), but only 9 have passed our $Age_{JX}$ selection criteria (\S \ref{methods.section}).  

Nevertheless, lines of independent evidence point towards the extreme youth of the cluster core. First, despite of the fact that the X-ray surveys tend to generate disk-free stellar samples \citep{Feigelson13}, all these nine $Age_{JX}$ sources are disk-bearing objects with optically thick disks. Seven of the nine have flat infrared spectral energy distributions, shown in Figure \ref{fig_seds}. Moreover, the apparent infrared-excess disk fraction for the 34 MPCMs with infrared-excess disk classification in the cluster core is extremely high, $94_{-7}^{+2}$\%.  Recalling that $Age_{JX}$ estimation does not use any information on infrared excesses, the high apparent disk fraction and flat infrared spectral energy distributions of the NGC~2024 core objects independent supports a very young age for the core region.

Second, the NGC~2024 cluster core is closely associated with the molecular core seen on the $Herschel$ image (Figure~\ref{fig1}a), suggesting that the cluster core is still embedded in its parental cloud. The median $J-H$ color for the $22$ MPCMs with available $JHK_s$ measurements in the NGC~2024 core is very high, $J-H = 3 \pm 0.1$~mag corresponding to $A_V \sim 20$~mag. Considering the integrated sample of 5525 MYStIX YSOs across 15 MYStIX star forming regions, G14 give a spline regression line relating $Age_{JX}$ and the $J-H$ absorption indicator can serve as an approximate age predictor for the most highly reddened clusters\footnote{Despite likely varying properties of the local molecular environments around the embedded MYStIX subclusters, the median $Age_{JX}$ estimates for these subclusters can be reasonably constrained. For instance, in Figure~4 of G14 showing the relationship between the $Age_{JX}$ and $J-H$, one can count over 70 YSOs from 10 different star forming regions within the $J-H$ color range between 2.7 and 3.3 mag. The median $Age_{JX}$ values for these YSOs is $0.6 \pm 0.1$~Myr (68\% confidence interval).}. According to this regression relationship, the age of the NGC~2024 core would be $\sim 0.6$~Myr.  

Third, the median X-ray column density for 95 MPCMs in the core with available $N_H$ estimates is remarkably high, $N_H = 5.7 \times 10^{22}$~cm$^{-2}$ corresponding $A_V \sim 35$~mag. This gives an age $\sim 0.1$~Myr from the spline regression relationship of G14, similar to the $\sim 0.2$~Myr value obtained above from the $Age_{JX}$ estimates directly (\S \ref{age_gradient_section}). 

These absorption measurements suggest that the entire MPCM sample in the core, not just the 9 stars with $Age_{JX}$ values, is extremely young. We finally note that since MYStIX and $Age_{JX}$ samples are incomplete for the  very young Class 0/I protostars (Feigelson et al. 2013; G14), the true median age could be even younger than our  $Age_{JX}$ estimates.

\subsection{Disk Fraction Evidence for the NGC~2024 and ONC Age Gradients \label{disk_fraction_section}}

For the $Age_{JX}$ stellar sample within the entire NGC~2024 cluster, all but a few stars have available disk classifications from $JHK+Spitzer$ infrared spectral energy distributions \citep{Povich2013}. These classifications are listed in Table~\ref{tbl_stars}. Figure \ref{fig_ks_vs_age}a plots the apparent disk fraction for the NGC~2024-$Age_{JX}$ stars as a function of median $Age_{JX}$ values in the elliptical annuli shown in Figure~\ref{fig1}. In addition to the clear jump from 100\% to $\sim 67$\% disk fraction from the core to the halo\footnote{The jump remains even when the $100$\% disk fraction estimate for the $Age_{JX}$ stellar sample within the cluster core is replaced by the $94$\% disk fraction estimate for the MPCM spectral energy distribution sample within the cluster core (\S \ref{ngc2024_core_section}).}, there is a hint of a decreasing disk fraction from 0.2~pc (green point) to $\sim 1$~pc (orange point) from the cluster center.  This is consistent with the gradient in $Age_{JX}$ values. Our estimates of the apparent disk fraction in NGC~2024 are roughly consistent with the estimate of \citet{HaischLadaLada2000} averaged over the entire cluster. Incidentally, we note that our apparent disk fraction of $\sim 60-70$\% for the $0.8-1.5$~Myr stars in the halo of NGC~2024 is similar to the disk fraction of $75$\% ($45$\%) for $1-2$~Myr solar-mass (low-mass) stars in the nearby Taurus star forming region \citep{Luhman2010}.

For the MPCM ONC catalog, \citet{Broos2013} employ the infrared spectral energy distribution disk classification results of \citet{Megeath2012}. However, there is a lack of disk classifications for the large number of MYStIX stars in the central part of the Orion region because of the extremely strong mid-infrared nebular background.  Therefore, here we use $K_s$-excess classification as a surrogate for disk fraction, understanding that this will result in systematically lower disk fractions than a classification using $JHK+Spitzer$ spectral energy distributions. A star is classified as a $K_s$-excess object if its position on the $JHK_s$ color-color diagram is to the right of the reddening line originating at $M=0.1$~M$_{\odot}$, assuming an age of $1$~Myr and the reddening law given in G14. All $Age_{JX}$ ONC stars have available $K_s$-excess measurements from the deep $JHK$ survey reported by \citet{Getman05} with results listed in Table~\ref{tbl_stars} here. Figure~\ref{fig_ks_vs_age}b shows that the $K_s$-excess frequency in the ONC core is significantly larger than that of the cluster periphery. Our estimate of 50\% disk fraction for the core agrees with the $50$\% $JHK$ excess frequency for the full Trapezium cluster derived by \citet{Lada2000}. Incidentally, we note that the $K_s$-excess frequency of $\sim 50$\% for the $\sim 1.2$~Myr stars in the ONC core is also consistent with the  $K_s$-excess frequency of $50$\% found for $\sim 0.8-1.5$~Myr stars in the W~40 region \citep{Kuhn2010,Getman2014}, and our $K_s$-excess frequency of $\sim 10-30$\% for the $\sim 1.9$~Myr stars in the halo of ONC is consistent with the  $K_s$-excess frequency of $20$\% found for $\sim 2-3$~Myr stars in the IC~348 region \citep{Lada1995}.  

Since optical (as well as mid-infrared) surveys suffer from the high background nebular emission in the central part of the ONC, the optical age catalog of \citet{DaRio2010} misses $\sim 70$\% [$\sim 60$\%] of the X-ray PMS members in the cluster core area [and the inner halo; green ellipse in Figure \ref{fig1}] in the mass range $(0.2-1.2)$~M$_{\odot}$.   \citet{Jeffries2011} find no correlations between optically derived isochronal ages and disk excess in the ONC; however, as their analysis is based on optical data, it has poor sensitivity towards the central part of the cluster where our current analysis detects significantly different ages and disk fractions.   

Since the universal $L_X - M$ relationship is the essence of the $Age_{JX}$ method (G14), we check that the discovered age gradients are not biased by possible differences in XLFs between disk-bearing and disk-free stars. When the core-halo annular strata from Figure \ref{fig1} are considered separately, we do not find any statistically significant differences between the XLFs in the disk-bearing and disk-free populations of NGC~2024 or the ONC. For all individual annular strata, the KS test probabilities range between $14$\% and $99$\%. When the $Age_{JX}$ subsamples are combined over all annular strata within the {\it Chandra} fields, we similarly find no differences between the XLFs in the ONC stars with and without $K_s$-excess ($P_{KS} = 47$\%; Figure \ref{fig_xlfs}d), consistent with the previous findings that PMS X-ray luminosities are not correlated with $K_s$ disk presence in the ONC \citep[see Figure 15 in][]{Preibisch2005}.  

But in NGC~2024, differences between XLFs of disk-bearing and disk-free stars may be present ($P_{KS} = 3.5$\%; Figure \ref{fig_xlfs}b). This may simply reflect an inhomogeneity in the NGC 2024 $Age_{JX}$ sample: due to the reduced 2MASS point source sensitivity in the central part of the NGC 2024 cluster, the $Age_{JX}$ sample appears to be missing many X-ray-weaker sources embedded in the molecular core (\S \ref{ngc2024_core_section}). However, as it is argued in \S \ref{ngc2024_core_section}, the ages of the missing population (or at least, of the X-ray detected fraction of it) are still expected to be consistent with our $Age_{JX}$ estimates for the cluster core, $\sim 0.2$~Myr, supporting the general finding of the core-halo age gradient in NGC 2024.

\section{Discussion} \label{discussion.section}

\subsection{Previous Evidence for Age Gradients}  \label{prev_evidence.section}

\subsubsection{Previous Reports on Stellar Ages in NGC 2024 and ONC}  \label{prev_evidence1.section}

Generally, the body of earlier work for the NGC~2024 and ONC clusters support our findings here, though they are perhaps individually less convincing than our results (\S \ref{results_section}). Our work has three strengths: (i) the underlying sample arises from a multiwavelength survey that efficiently finds both disk-bearing and disk-free stars \citep{Feigelson13}; (ii) the underlying age estimator applies to both disk-bearing and disk-free stars and gives persuasive age large-scale gradients in many star forming regions (G14); and (iii) the sample studied is limited to a narrow range of stellar masses around $0.2-1.2~M_\odot$ so there is no conflation of mass segregation and age gradient effects.

\begin{description}

\item[NGC 2024 cluster] There have been no previous reports of a spatial-age gradient in the stellar cluster that ionizes the Flame Nebula.  Studies have reported a very young average age for the cluster.  \citet{Ali98} perform near-infrared photometry and spectroscopy on 40 members and report ``that the cluster is fairly young (average age ~0.5 Myr) and coeval''.   A similar study with a larger sample of $\sim 70$ cluster members arrives at a similar median cluster age (0.5~Myr) but with a wide age spread from $<<$1 to $\sim 30$~Myr \citep{Levine06}.  These studies cannot reveal a cluster-wide age gradient as their near-infrared samples miss most of the members in the inner regions due to high obscuration ($A_V \geq 20$~mag) and high background nebular emission.  \citet{Haisch01} report a very high fraction of stars with infrared-excess disks (85\% of 59 stars) and infer a very young age around $\sim 0.3$~Myr.  However, their underlying sample misses most disk-free stars that are found in X-ray studies, so their disk fraction could be overestimated (Figure \ref{fig_ks_vs_age}a). \citet{Burgh12} discuss how the far-ultraviolet spectrum of NGC~2024 IRS~1 has strong P Cygni wind feature characteristics of a $\sim 3$~Myr old O8 I supergiant, rather than a younger late-O main sequence star.  But the finding is difficult to interpret and may be due to scattered light from an obscured star behind the dust lane of the nebula. 

\item[Orion Nebula Cluster] From an HR diagram constructed from {\it Hubble Space Telescope} photometry and ground-based photometry and spectroscopy, \citet{Reggiani2011} calculate ages of $\sim$1000 stars in the ONC based on \citet{Siess2000} evolutionary models.  Their analysis is limited to masses above $\sim 0.4$~M$_\odot$ where the sample is $>80$\% complete.  The age distribution over most of the cluster is approximately constant with mean age around 2.2~Myr and an overall age spread of $2-3$~Myr.  In their spatial analysis of the age distribution, Reggiani et al. obtain a result similar to that reported here (\S\ref{results_section}): most of the cluster volume has a constant age distribution, but the inner core region (radius $\sim 0.2$~pc) has an excess of younger stars with ages $<1.2$~Myr.  However, they caution that this result could be affected by a bias in the completeness of the sample, as their optical sample may be less complete near the cluster center where the nebular background emission is brighter. They conclude with the statement, ``if this `age segregation' is confirmed, this may have implications for the study of the dynamical evolution of the cluster''. Unlike, the optical sample of Reggiani et al., the $Age_{JX}$ sample was built to be nearly complete and uniform in $L_X$, and thus in mass, across the entire ONC cluster within the {\it Chandra} field (\S \ref{xlf_section}).

\end{description}

\subsubsection{Reports on Age Gradients in Other Regions}  \label{prev_evidence2.section}

A number of earlier studies in recent years have results relating to the age gradients presented here.  We review these here.

\begin{description}

\item[Serpens North, Serpens South, and Corona Australis] Serpens North \citep{Gorlova2010}, Serpens South \citep{Gutermuth2008},  and Corona Australis \citep{Peterson2011} are nearby ($D < 400$~pc) star formation regions with stellar clusters less rich than those studied with MYStIX ($\sim 50$ observed disk-bearing objects per cluster). Their morphologies, with primarily protostellar cores and PMS halos, are reminiscent of that of NGC 2024. By comparing the observed ratios of protostars and Class II PMS stars in these clusters to the predictions of his analytical cluster model of ``constant birthrate, core-clump accretion, and equally likely stopping" \citet{Myers2012} reports the core-halo age gradients from $\sim 0.3$~Myr to $\ga 1$~Myr. Myers speculates that the astrophysical cause of these gradients could be due to later formation of dense central parts in parental molecular filaments, similar to our scenarios (a) and (b) below (\S \ref{scenarios_section}).

\item[W 3 Main] The rich embedded cluster W 3 Main within the giant W~3/W~4/W~5 molecular cloud complex has a unique collection of radio H\,{\sc{ii}} regions from embedded OB stars within a $\sim 0.3$~pc radius core \citep{Tieftrunk98}.  These ionized bubbles have a range of sizes, but some are very small ultracompact and hypercompact H\,{\sc{ii}} regions implying  that the related massive stars are extremely young with ages $\sim 10^4$~yr.  In an X-ray/infrared study similar to, but preceding, the MYStIX project, \citet{Feigelson08} identify several hundred PMS stars tracing a large spherical cluster with radius $\sim 2$~pc.  Only a very small fraction of the PMS population have infrared-excess protoplanetary disks, implying that the extended cluster has age $\geq 2$~Myr, much older than the central OB stars. \citet{Bik13} confirm this result and extend it with suggestive evidence from spatial analysis of $K$-band excess stars and the $K$ luminosity function that the younger PMS stars, like the OB stars, are  centrally concentrated.  Altogether, it appears that W~3 Main is similar to NGC~2024 and the ONC with a spatial-age gradient in a rich young star cluster.

\item[IRAS 19343+2026]  \citet{Ojha10} study the stars in the embedded cluster IRAS~19343+2026 at a distance around 4~kpc.  Here several dozen PMS stars with estimated ages $\geq 1$~Myr lie in a region about 1~pc across around a core of several mid-B stars with strong infrared-excess disks.  The latter have estimated ages around $\sim 10^5$~yr, younger than the more extended PMS cluster.    

\item[Bright rimmed clouds]  Several cases have been found of small-scale sequential star formation in PMS stars on the periphery of rich clusters where the H\,{\sc{ii}} region shocks the ambient molecular material and produces ionized bright rimmed clouds (BRCs).  In these cases, the older stars lie closer to the central OB-dominated cluster and the younger stars lie embedded in the molecular cloud.  This phenomenon is seen in near-infrared samples associated with BRC 11NE, BRC 13, BRC 14, BRC 37 among others \citep{Ogura07, Chauhan09, Chauhan11} and in X-ray samples associated with IC 1396N, Cepheus B, and IC 1396A \citep{Getman07, Getman09, Getman12}.  A general trend in these and other regions \citep[e.g., the Rosette Nebula;][]{Wang10} is found that PMS stars on the periphery of giant H\,{\sc{ii}} regions are younger than the PMS stars in the central rich cluster.  These spatial-age sequences are attributed to star formation in the surrounding cloud triggered by the expansion of the H\,{\sc{ii}} region over millions of years, plausibly through some combination of `radiative driven implosion' and `collect and collapse' mechanisms.  The results are age gradients opposite to those discussed here with older stars at the cluster core and younger stars at the cluster periphery.  These findings relating to secondary triggering processes do not provide insight into the formation process of the original rich cluster that produced the giant H\,{\sc{ii}} region.

\end{description}

\subsection{Scenarios for Intracluster Age Gradients \label{scenarios_section}}

We establish that the median age of low mass pre-main sequence stars in cluster cores is younger than the median age of similar stars in cluster peripheries in the two richest star clusters of the Orion molecular cloud.  Our results give no information in any direction on the sequence of events giving rise to OB star formation or mass segregation.  We also do not provide clear results on the age spread, either globally or in spatially confined regions.  For example, Figure~\ref{fig1} (panels b and e) suggests that a possible cause of the gradients in median may be mostly due to the absence of older stars, not an excess of younger stars, in these clusters.

The simplest model for the spatial-age gradient we see in the NGC~2024 and Orion Nebulae clusters is that stars form first in the periphery of the progenitor molecular core and later in the center.  But this is astrophysically counterintuitive.  The classical scenario for unified cluster formation predicts that the highest density core at the center of a cloud concentration collapses first, while star formation in the cluster outskirts occurs at different times \citep[e.g.,][]{Klessen11}. \citet{Parmentier2013} find that the star-formation rate decelerates faster in regions of higher density as gas is quickly exhausted on the shorter free-fall timescale, implying that the dense centers of molecular clouds should form most of their stars early in the cluster formation process. \citet{Pfalzner11} argues that a star-formation front moving from the inside to the outside of a cloud can produce a relationship between cluster mass and radius that she observes in a sample of rich clusters.  These scenarios would suggest a radial age gradient opposite from what is observed in our work: the cluster centers would have the oldest stars, not the youngest stars as we find here.

The properties of the natal molecular cloud must be taken into consideration when determining what scenarios are plausible explanations for our effect. Clouds in massive star forming regions (MSFRs) have large scale density gradients, are turbulent \citep{MacLow04} with dense, gravitationally unstable filaments \citep{Andre10, Hacar13}.  Gas probably flows perpendicularly onto the filaments from the larger cloud, and longitudinally along the filaments onto star-forming clumps \citep{Hacar11, Palmeirim13, Gomez13, Peretto13b}.  Gas is later expelled by feedback effects of young stars including protostellar outflows \citep[e.g.][]{Li06}, OB radiation and winds \citep[e.g.][]{Dale08,Dale13,Townsley11}, and supernovae \citep{Dekel13}.

Star inflow, as well as gas inflow, may be important. In a model where star formation is distributed throughout the molecular cloud in local turbulent concentrations of gas, these stars are gravitationally pulled inward to form the cluster core through violent relaxation \citep{McMillan07,Bate09}\footnote{The reader is encouraged to view the video accompanying the simulation of a cluster forming in a turbulent cloud by \citet{Bate09} available at \url{http://www.astro.ex.ac.uk/people/mbate/Cluster/cluster3d.html}.}.

Models of cluster formation do not necessarily require delayed star formation in the core to produce the spatial-age gradient in lower mass stars reported here. Instead, older stars could be removed from the core or star formation could continue in the core after it has terminated in the outer regions. Recall that we only report that the median age in the cluster cores is lower than in the cluster halos; we do not establish the full distribution of ages in either location, and these distributions need not have similar shapes. 

Logically, there are three possible causes for the age gradient that is observed in the current work: (1) star-formation occurred more recently in the inner cluster region than the outer region; (2) older stars move outward; or (3) younger stars move inward. Any combination of these three processes would also produce an age gradient. Below we review star-formation theory that might lead to these phenomena.  Three of these scenarios are illustrated by the cartoons in Figure \ref{cartoon.fig}. The panels -- relating to scenarios A, F, and G -- differ in their assumptions of the global structure of the molecular cloud, filamentary cloud substructure, and infalling cloud clumps.

\subsubsection{More recent star formation in the cluster center} 

A. Although gas accretion onto stars will deplete the cloud material in both regions, gas is less dense in the periphery. There is likely a threshold below which gas is no longer unstable to gravitational collapse. \citet{Andre10} find that this threshold is roughly a column density of $N_{H_2} \sim 7\times10^{21}$~cm$^{-2}$, corresponding to $A_V \simeq 8$~mag extinction, in dense filaments that pervade many MSFRs. With decreasing gas surface density throughout the cloud, the less dense peripheral cloud material will drop below this threshold first and stop forming stars before the cloud center stops.  Viewed later as a MYStIX star forming region, the outer stars will appear older on average than the central stars.  

B. \citet{Palla00} and \citet{Stahler06} argue that over millions of years the combination of global gravitational contraction and star formation threshold results in acceleration of the star formation rate. Thus, most of the stars in the core would have formed at the end of the cloud lifetime and would appear younger than stars in the cluster periphery.

C. Many theoretical models of massive star formation predict that O stars form at the center of clusters \citep{Zinnecker07}. Furthermore, in a few clusters observational evidence indicates that O stars may be the last stars in a cluster to form \citep{Ojha10,Feigelson08}, perhaps because the timescale needed for competitive accretion or mergers to make massive stars is longer. This would imply, that -- at least for massive stars -- there would be age stratification. 

\subsubsection{Older stars move outward}

D. If stars are forming from a molecular core that is not gravitationally bound, then the resulting cluster may be supervirial and its stars will drift away from the cloud after they are born due to their initial velocities. This may be common for low-mass cloudlets in regions like the South Pillars of Carina \citep{Povich11}. Unbound stars could also be produced by three-body interactions; for example, star-cluster formation simulations by \citet{Bate09} shows ejection of some first-generation stars. Older stars will have a longer time to drift, so they will be located farther from the cluster center. After some first-generation of stars has drifted away, infall of cloud material can lead to a new population of young stars being born at the cluster center.   

E. There is growing evidence to suggest that stars are born with subvirial velocity dispersions.  In hydrodynamical models, \citet{Bate03}  find that initial stellar velocities are subvirial even when the simulated gas cloud start unbound.  Models of subvirial clusters match radial velocity observations of the Orion star-forming region \citep{Proszkow09}. In this case, stars would tend to get dynamically heated with cluster relaxation. Since older stars would have more time to experience dynamical effects, they would preferentially be located at larger radii than younger, dynamically cooler stars.

F.  Several phenomena can cause young cluster expansion.  Most important is the loss of the gravitational potential of the molecular material which dominates the early phases \citep[e.g.,][]{Lada84, Moeckel12}.  Stellar dynamics can cause expansion through three-body interactions with hard binaries and gravothermal core collapse associated with mass segregation \citep{Giersz96, Pfalzner13}. MSFRs often have multiple subclusters of young stars related to the clumpy structure of the natal molecular cloud \citep[e.g.,][]{Elmegreen00,Kuhn2014}. Thus, the first-born subclusters may have already expanded while the later-born subclusters are still highly centrally concentrated. The expanded older subclusters may resemble a low surface-density halo of stars, while the younger, centrally concentrated subclusters may form the core of the resulting cluster.

\subsubsection{Younger stars move inward}

G. Some simulations of cluster formation show star-forming filaments of gas falling into the gravitational potential minimum at the center of a MSFR, bringing along many of the young stars that formed in the filaments \citep[e.g.][]{Bate03,Bate09}. These filaments of stars and gas are more massive than unembedded stars making up a star cluster at the center of a star-forming region, so they will sink to the center due to dynamical segregation \citep[e.g.,][]{Clarke10}. If stars in these star-forming filaments are younger than the rest of the cluster, this can lead to the observed age stratification.

H. If subclusters expand as they age, younger subclusters are likely to be smaller and denser \citep{Pfalzner09}. Simulations of subcluster mergers indicate that the resulting cluster may generate a core-halo structure \citep{Bate09}, which may include stellar age stratification. If an older low-density subcluster merges with a younger high-density subcluster, the older stars will likely become more widely distributed.

\section{Directions of further research}

The findings here involving two clusters in the nearby Orion Molecular Cloud complex, combined with past studies of W~3 Main and IRAS 19343+2026, could have important general implications for the formation processes of clustered star formation.  But further research to establish spatial-age gradients in a larger cluster sample and astrophysical modeling to discriminate between the models above (or other explanations) is highly desirable.
  
We see a clear path towards observational elucidation of intra-cluster spatial-age gradients in rich clusters.  Obtaining deeper $Chandra$ and/or near-infrared imagery for MYStIX and similar cluster targets is an important ingredient.  The Orion Nebula region has already been the subject of a very long observation \citep[COUP;][]{Getman05}. Improved NIR (currently only 2MASS) and X-ray (currently  80~ks long) exposures of NGC~2024 would increase the stellar sample with stars in the $0.1-0.5$~M$_\odot$ range by a factor of two or more. Deeper exposures for carefully selected clusters beyond the Orion molecular cloud would give sufficiently rich stellar samples to allow detection of spatial-age gradients, if present. For many nearby star forming regions, existing $JHK$ and mid-infrared samples are already sufficiently deep for finding counterparts to the fainter X-ray sources.

Hydrodynamical and N-body simulations of different modes of cluster formation and early evolution, such as those illustrated in Figure~\ref{cartoon.fig}, could elucidate the origins of our spatial-age segregation.  Several research groups are developing relevant models including \citet{Bate09}, \citet{Allison11}, \citet{Moeckel12}, \citet{Parker12}, \citet{Pfalzner13} and \citet{Banerjee13}. The MYStIX project not only provides positions and mass estimates of individual cluster members, but also measures of the total stellar population, central star density, and other parameters relevant to astrophysical modeling (Kuhn et al., in preparation).  Theoretical studies might not only constrain the modes of cluster formation, but also the epoch of gas expulsion that is crucial for the dynamical expansion and dissipation of clusters.

\section{Conclusions} \label{conclusion_section}

It has not been clearly known whether stellar cluster formation occurs within 1-few core free-fall times or over millions of years. The view of the slow mode of cluster formation is supported in part by empirical evidence of wide age spreads within rich stellar clusters. However, it has been unclear whether these age spreads are due to poor individual age estimates or to real extended star formation histories (\S \ref{intro_section}).

In this work, we analyze age distributions of two nearby rich stellar clusters, the NGC 2024 (Flame Nebula) and Orion Nebula clusters in the Orion molecular cloud complex. Our analysis is based on a new estimator of PMS stellar ages, $Age_{JX}$, that is derived from X-ray and near-infrared photometric data (G14). To overcome the problem of uncertain individual ages, we compute median ages and their confidence intervals of stellar samples within annular subregions of the clusters. We find core-halo age gradients in both the NGC 2024 and Orion Nebula clusters: low-mass PMS stars in cluster cores appear younger, and thus formed later, than low-mass PMS stars in cluster peripheries (\S \ref{age_gradient_section}). Various selection effects were investigated (\S \ref{xlf_section}). In the ONC they turned out to be unimportant, but not in NGC 2024. Other considerations helped suggesting that these should not be a major bias in the result (\S \ref{ngc2024_core_section}). These findings are further supported by the independent discovery of the spatial gradients in the mid-infrared and near-infrared disk fractions (\S \ref{disk_fraction_section}). Based purely on PMS stellar data in a narrow mass range, our results give no information on the sequence of events that give rise to OB star formation and mass segregation.

\begin{figure}
\centering
\includegraphics[angle=0.,width=6.5in]{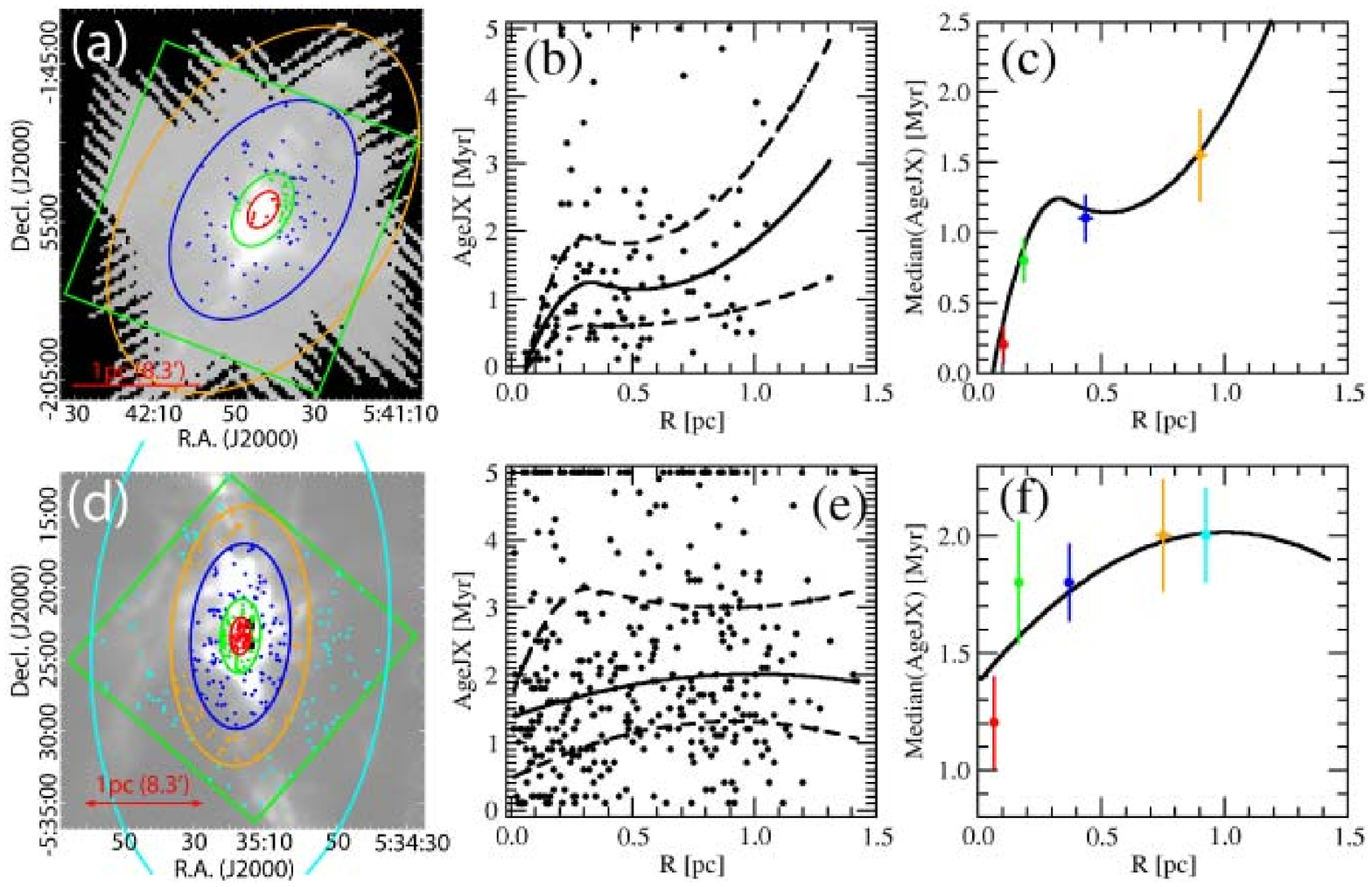}
\caption{\small Age analysis within the NGC~2024 (top row) and Orion Nebula (bottom row) clusters.  Panels a and d show MYStIX-$Age_{JX}$ stars superimposed on the $500\mu$m {\it Herschel}-SPIRE images. The stars are stratified in elliptical annular regions.  The {\it Chandra}-ACIS fields defining the MYStIX samples are outlined in green. Panels b and e show  $Age_{JX}$ as a function of angular distance from the centers of the clusters for individual $Age_{JX}$ stars. The solid (dashed) lines show the relationship of the 25\%, 50\% (median), and 75\% quartiles of $Age_{JX}$ and distance obtained from B-spline regression. Panels c and f show $Age_{JX}$ as a function of angular distance from the cluster centers for the region-stratified subsamples drawn in panels a and d.  For each subsample, its median($Age_{JX}$) and 68\% confidence interval on that median are marked by a color point and an error bar, respectively. The median regression lines from panels (b and e) are over-plotted as black curves. \label{fig1}}
\end{figure}

\begin{figure}
\centering
\includegraphics[angle=0.,width=6.in]{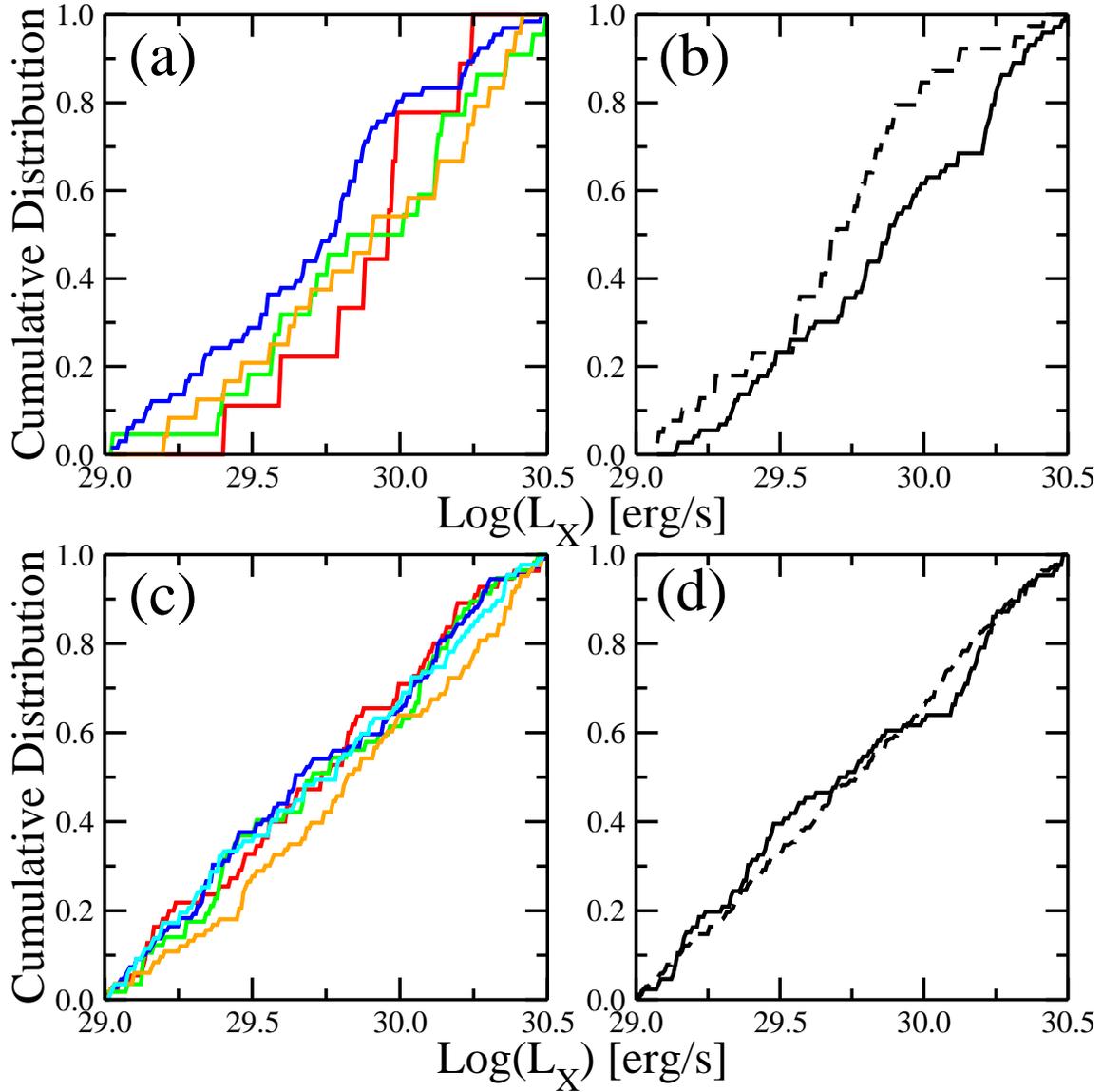}
\caption{Cumulative distributions of X-ray luminosity functions for the NGC~2024 (panels a, b) and ONC (panels c, d) $Age_{JX}$ stars. Panels (a, c) present XLFs stratified and color-coded by the annular regions that are shown in Figure~\ref{fig1}. Panels (b, d) present XLFs for the stars across the entire clusters that are stratified by the presence/absence of disks. The XLFs of the NGC~2264 disk-bearing and ONC $K_s$-excess objects are represented by the solid lines; the XLFs of the NGC~2264 disk-free and ONC non-$K_s$-excess objects are represented by the dashed lines. \label{fig_xlfs}}
\end{figure}

\begin{figure}
\centering
\includegraphics[angle=0.,width=7.0in]{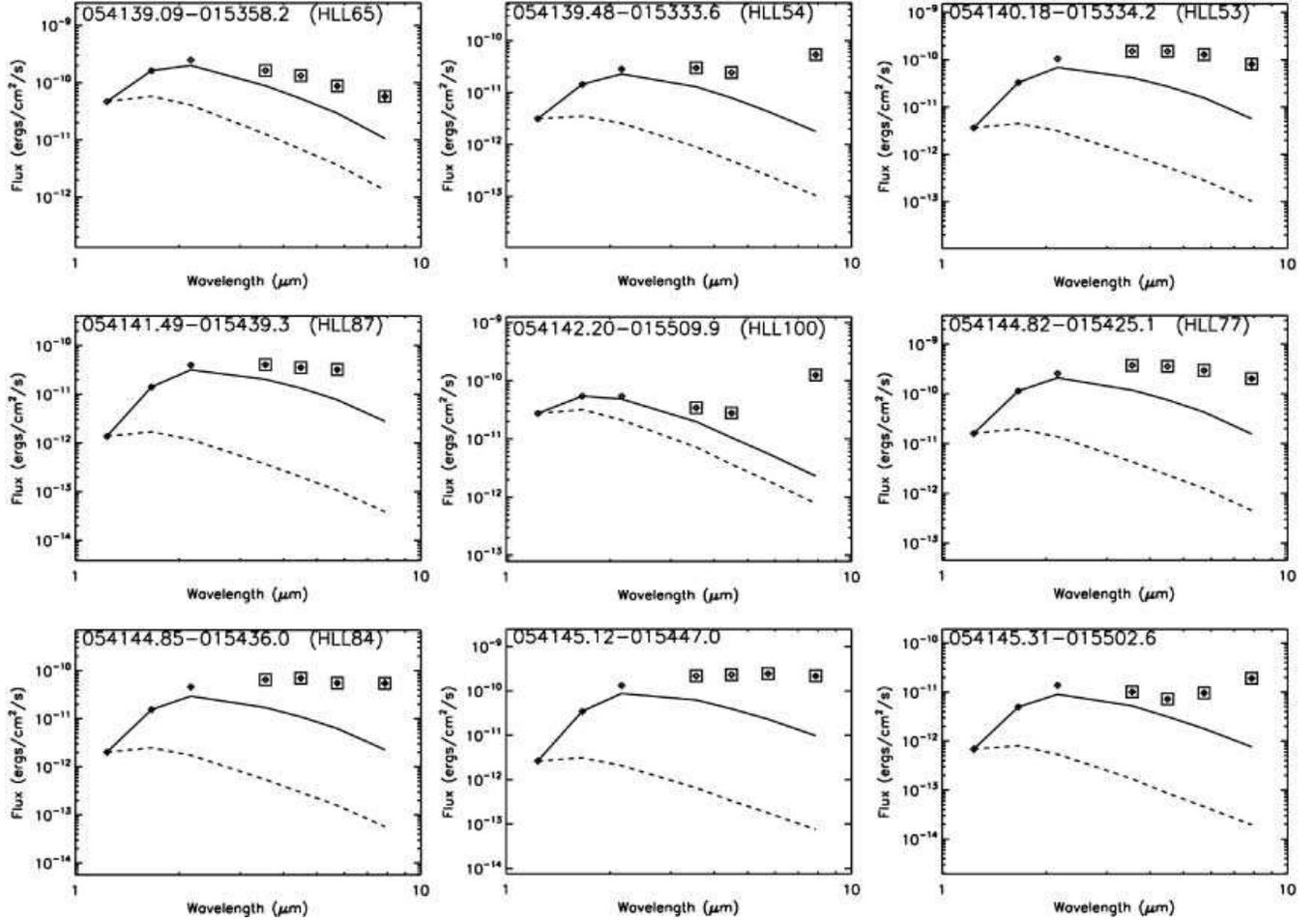}
\caption{IR SEDs for the 9 $Age_{JX}$ sources in the core of the NGC~2024 cluster (red in Figure~\ref{fig1}). The dashed and solid lines give the apparent-original and (de)reddened median template SEDs from \citet{Lada2006} fitted to the data. The figure legend gives the MYStIX source designation, and in parenthesis, if available, the IR counterpart from \citet{HaischLadaLada2000}. \label{fig_seds}}
\end{figure}

\begin{figure}
\centering
\includegraphics[angle=0.,width=6.5in]{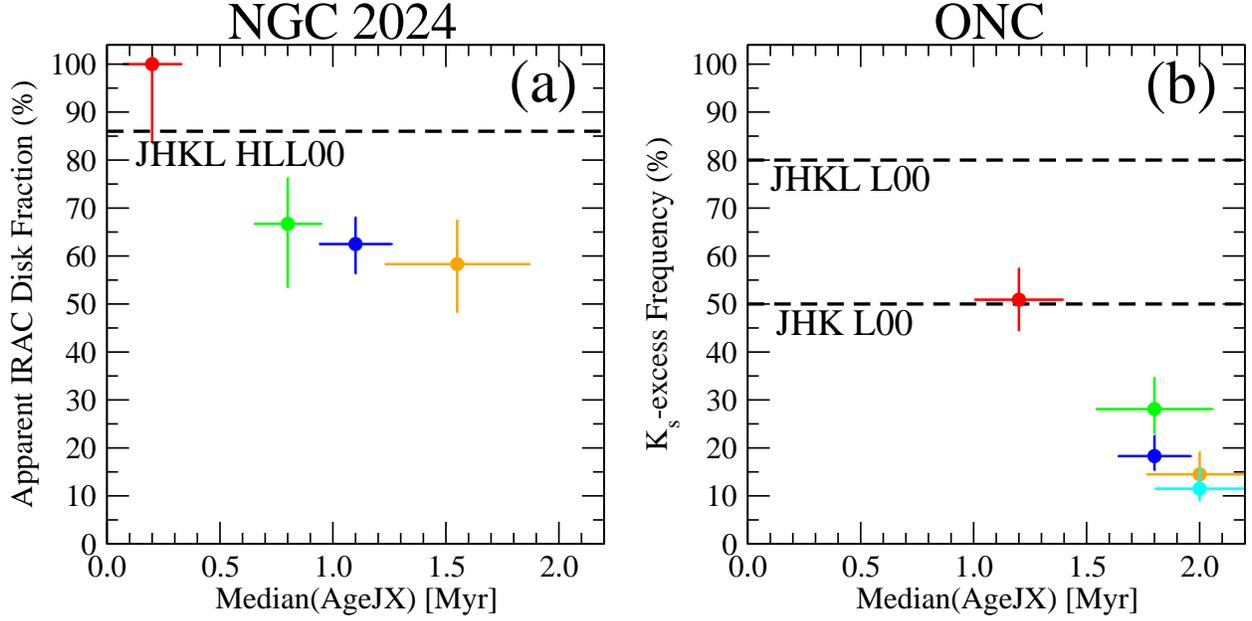}
\caption{Panel (a) presents IRAC Disk Fraction for the NGC 2024-$Age_{JX}$ stars as a function of age. The dashed line indicates the $86$\% $JHKL$ excess/disk fraction from \citet{HaischLadaLada2000}. Panel (b) presents $K_s$-excess frequency for the ONC-$Age_{JX}$ stars as a function of age. The dashed lines indicate $80$\% $JHKL$ excess/disk fraction and $50$\% $JHK$ excess frequency for the Trapezium cluster reported in \citet{Lada2000}. On both panels, the different color points correspond to the stellar subsamples shown in Figure~\ref{fig1}. The error bars were calculated using binomial distribution statistics described in \citet{Burgasser2003}. \label{fig_ks_vs_age}}
\end{figure}

\begin{figure}
\centering
\includegraphics[angle=0.,width=6.0in]{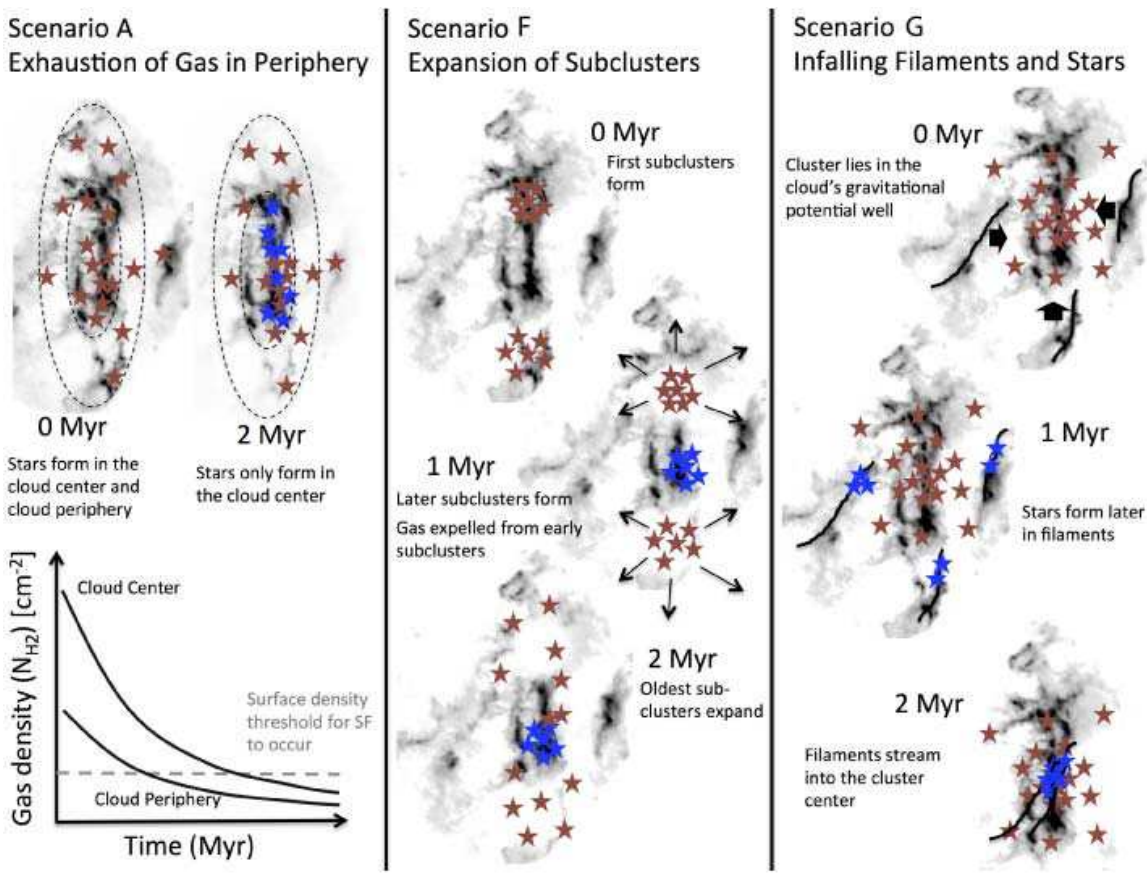}
\caption{The three panels show cartoons depicting different paths to radial age stratification in MSFRs (Scenarios A, F, and G, from left to right, respectively). The young cluster members are shown by the brown (earlier forming) and blue (later forming) stars. A representative cloud is shown in gray scale (adapted from {\it Herschel} images of IC~5146), which was modified to show a gas surface-density gradient (left), gas expulsion (middle), and infalling filaments (right). The graph (left panel) shows gas surface density versus time for the inner and outer cloud, as it falls below a critical threshold for star formation. The text annotations explain the part of the scenario occurring in each at various age steps over a period of $\sim$2~Myr, which result in the younger (blue) cluster members lying near the center of the region than the older (brown) cluster members lying in the periphery of the region.
\label{cartoon.fig}}
\end{figure}

{\it An essential implication of our findings is that clusters do have true age spreads.} If clusters truly formed exceedingly rapidly, then our median $Age_{JX}$ estimates would be consistent with a single cluster age value. The ages shown in Figure~\ref{fig1} (panels c and f) invalidate rapid cluster formation for NGC~2024 and the ONC. 

We provide several theoretical star formation scenarios that alone, or in combination with each other, may lead to the observed core-halo age gradients (\S \ref{scenarios_section}). These include: exhaustion of gas in the periphery of a cloud, acceleration of star formation in the cloud core, late formation of massive stars with low-mass stellar siblings, kinematic outward drift of older stars, stellar dynamical heating with cluster relaxation, expansion of subclusters, infalling filaments nursing newly born stars, and subcluster mergers.

\acknowledgements

{\bf Acknowledgements:} ~~ We thank P.~Broos (Penn State), L.~Townsley (Penn State), T. Naylor (University of Exeter), M. Povich (California State Polytechnic University), and K. Luhman (Penn State) for development of X-ray and IR analysis tools, for participation in production of MYStIX catalogs, for their expert assistance with the Flame and Orion data, and for their help in the development of the $Age_{JX}$ estimator. We also thank the anonymous referee for his time and
useful comments that improved this work. The MYStIX project is supported at Penn State by NASA grant NNX09AC74G, NSF grant AST-0908038, and the $Chandra$ ACIS Team contract SV4-74018 (G.~Garmire \& L.~Townsley, Principal Investigators), issued by the $Chandra$ X-ray Center, which is operated by the Smithsonian Astrophysical Observatory for and on behalf of NASA under contract NAS8-03060. The Guaranteed Time Observations (GTO) included here were selected by the ACIS Instrument Principal Investigator, Gordon P. Garmire, of the Huntingdon Institute for X-ray Astronomy, LLC, which is under contract to the Smithsonian Astrophysical Observatory; Contract SV2-82024. This research made use of data products from the $Chandra$ Data Archive and the {\it Spitzer Space Telescope}, which is operated by the Jet Propulsion Laboratory (California Institute of Technology) under a contract with NASA. This research used data products from the Two Micron All Sky Survey, which is a joint project of the University of Massachusetts and the Infrared Processing and Analysis Center/California Institute of Technology, funded by the National Aeronautics and Space Administration and the National Science Foundation.

\clearpage
\newpage

\begin{deluxetable}{ccccccccc}
\centering \tabletypesize{\footnotesize} \rotate \tablewidth{0pt} \tablecolumns{9}

\tablecaption{$Age_{JX}$ Estimates for NGC 2024 and ONC Stars \label{tbl_stars}}

\tablehead{\colhead{Cluster} & \colhead{Source} & \colhead{R.A.} & \colhead{Decl.} & \colhead{$R$} & \colhead{$\log(L_X)$} & \colhead{$Age_{JX}$} & \colhead{Disk} & \colhead{Annulus}\\

\colhead{} & \colhead{MPCM} & \multicolumn{2}{c}{(J2000 deg)} & \colhead{(pc)} & \colhead{(erg s$^{-1}$)} & \colhead{(Myr)} & \colhead{Flag} & \colhead{Flag}\\

\colhead{(1)} & \colhead{(2)} & \colhead{(3)} & \colhead{(4)} & \colhead{(5)} & \colhead{(6)}  & \colhead{(7)}  & \colhead{(8)}  & \colhead{(9)}}

\startdata
NGC 2024 & 054117.99-014929.2 &    85.324989 &    -1.824787 & 0.93 & 29.91 & 0.5 & DSK & [6-10]\\
NGC 2024 & 054118.38-014953.1 &    85.326618 &    -1.831437 & 0.89 & 30.31 & 1.3 & DSK & [6-10]\\
NGC 2024 & 054119.18-015226.8 &    85.329942 &    -1.874130 & 0.73 & 29.28 & 1.1 & NOD & [2-6]\\
NGC 2024 & 054120.68-015858.2 &    85.336183 &    -1.982844 & 0.87 & 30.12 & 4.7 & DSK & [6-10]\\
NGC 2024 & 054125.01-015226.7 &    85.354241 &    -1.874088 & 0.57 & 29.90 & 1.6 & NOD & [2-6]\\
NGC 2024 & 054125.53-014809.2 &    85.356376 &    -1.802560 & 0.89 & 29.78 & 0.8 & DSK & [2-6]\\
NGC 2024 & 054125.71-015545.9 &    85.357128 &    -1.929424 & 0.54 & 29.34 & 0.7 & DSK & [2-6]\\
NGC 2024 & 054125.88-015728.6 &    85.357855 &    -1.957963 & 0.63 & 29.41 & 1.2 & NOD & [6-10]\\
NGC 2024 & 054126.11-020016.1 &    85.358804 &    -2.004496 & 0.88 & 30.37 & 0.6 & DSK & [6-10]\\
NGC 2024 & 054126.94-015451.7 &    85.362263 &    -1.914380 & 0.47 & 29.28 & 1.1 & NOD & [2-6]\\
\enddata

\tablecomments{This table is available in its entirety (121 NGC~2024 and 391 ONC stars) in the machine-readable form in the on-line journal. A portion is shown here for guidance regarding its form and content. For the information on numerous other stellar quantities, this table can be linked to Table 2 of \citet{Broos2013} using the source name keyword (Column 2). Column 1: Name of a cluster. Column 2: MYStIX source's IAU designation. Columns 3 and 4: Right ascension and declination for epoch J2000.0 in degrees. Column 5: Angular distance from the center of the cluster in pc. Column 6:  Intrinsic X-ray luminosity in the $(0.5-8)$~keV band from \citet{Broos2013}. Column 7: $Age_{JX}$ estimate from G14. Individual age estimates of 5~Myr indicate that ages run into the truncation limit of 5~Myr (G14). Column 8: Disk flag. For NGC~2024: 'DSK' and 'NOD' - disk-bearing and disk-free young stellar objects; the classification is adopted from the MIRES catalog of \citet{Povich2013} based on NIR and $Spitzer$ IRAC photometry.  For ONC: 'Ks' and 'noKs' - PMS stars with and without $K_s$-excess (\S \ref{disk_fraction_section}); the classification is adopted from \citet{Getman05} based on NIR photometry. Column 9: Annular region stratum in units of core radii from Figure \ref{fig1}.}
\end{deluxetable}

\end{document}